\documentclass[12pt,onecolumn]{IEEEtran}
\usepackage{amsmath,amsthm,amssymb}
\usepackage[pdftex]{graphicx}
\graphicspath{{./fig/}}
\DeclareGraphicsExtensions{.pdf,.jpeg,.png}
\usepackage{hyperref}

\providecommand{\corollaryname}{Corollary}
\providecommand{\definitionname}{Definition}
\providecommand{\lemmaname}{Lemma}
\providecommand{\remarkname}{Remark}
\providecommand{\theoremname}{Theorem}

\theoremstyle{remark}
\newtheorem{thm}{\protect\theoremname}
\newtheorem{defn}{\protect\definitionname}
\newtheorem{lem}{\protect\lemmaname}
\newtheorem{rem}{\protect\remarkname}
\newtheorem{cor}{\protect\corollaryname}

\newcommand{\pch}{p_{\text{ch}}}
\newcommand{\lw}{\lambda_{\text{w}}}
\newcommand{\Aw}{A_{\text{w}}}
\newcommand{\Ac}{A_{\text{c}}}
\newcommand{\An}{A_{\text{nodes}}}
\newcommand{\Cc}{C_{\text{circuit}}}
\newcommand{\Cu}{C_{\text{unit-area}}}
\newcommand{\Vdd}{V_{\text{DD}}}
\newcommand{\tech}{\xi_{\text{tech}}}
\newcommand{\Ktech}{K_{\text{tech}}}
\newcommand{\Eproc}{E_{\text{proc}}}
\newcommand{\Edec}{E_{\text{dec}}}

\newcommand{\half}{\frac{1}{2}}
\newcommand{\fifth}{\frac{1}{5}}

\begin{document}

\title{Energy Consumption of VLSI Decoders\thanks{\noindent Submitted
for publication on November 7th, 2013, revised November 28th, 2014.  Presented in part
at the 2013 Canadian Workshop on Information Theory,
June 18--21, Toronto, Canada, 2013.}}

\author{Christopher Blake and Frank R. Kschischang\\Department of
Electrical \& Computer Engineering\\University of Toronto\\
\texttt{\small christopher.blake@mail.utoronto.ca} \texttt{\small frank@comm.utoronto.ca}}
\maketitle
\begin{abstract}
Thompson's model of VLSI computation relates the energy of a computation to
the product of the circuit area and the number of clock cycles needed to
carry out the computation. It is shown that for any family of circuits
implemented according to this model, using any algorithm that performs
decoding of a codeword passed through a binary erasure channel, as the
block length approaches infinity either (a) the probability of block error
is asymptotically lower bounded by $\half$ or (b) the energy of the
computation scales at least as $\Omega\left(n\sqrt{\log n}\right)$, and so
the energy of successful decoding, per decoded bit, must scale at least as
$\Omega\left(\sqrt{\log n}\right)$. This implies that the average energy
per decoded bit must approach infinity for any sequence of codes that
approaches capacity. The analysis techniques used are then extended to the
case of serial computation, showing that if a circuit is restricted to
serial computation, then as block length approaches infinity, either the
block error probability is lower bounded by $\half$ or the energy scales at
least as fast as $\Omega\left(n\log n\right)$. In a very general case that
allows for the number of output pins to vary with block length, it is shown
that the average energy per decoded bit must scale as
$\Omega\left(n\left(\log n\right)^{\fifth}\right)$. A simple example is
provided of a class of circuits performing low-density parity-check
decoding whose energy complexity scales as $O\left(n^2 \log \log n\right)$.
\end{abstract}

\section{Introduction}
 
\IEEEPARstart{S}{ince} the work of Shannon \cite{Shannon}, information
theory has sought to determine how much information can be communicated
over a noisy channel; modern coding theory has sought ways to achieve this
capacity using error control codes. A standard channel model is the
additive white Gaussian noise (AWGN) channel, for which the maximum rate of
information that can be reliably communicated (known as the capacity) is
known and depends on the transmission power. This model does not, however,
consider the energy it takes to encode and decode; a full understanding of
energy use in a communication system requires taking into account these
encoding and decoding energies, along with the transmission energy.
Currently there has been very little work done in seeking a fundamental
understanding of the energy required to execute an error control coding
algorithm.

Early work in relating the area of circuits and number of clock cycles in
circuits that perform decoding algorithms was presented by El Gamal
\emph{et al.} in \cite{ElGamal}. More recent work in trying to find
fundamental limits on the energy of decoding can be attributed to Grover
\emph{et al.} in \cite{groverFundamental}.  In this work, the authors
consider decoding schemes that are implemented using a VLSI model
attributed to Thompson \cite{ThompsonThesis} (which we will describe
later), and they are able to show that for any code, using any decoding
algorithm, as the required block error probability approaches $0$, the sum
of the transmission, encoding, and decoding energy, per bit, must approach
infinity at a rate of $\Omega\left(\sqrt[3]{\frac{1}{\log
P_{e}^{\text{blk}}}}\right)$, where $P_{e}^{\text{blk}}$ is the block error
probability of the code.  This result is useful to the extent that it
suggests how to judge the energy complexity of low error probability codes;
however, it does not suggest how the energy complexity of decoding scales
as capacity is approached.

The result of this paper uses a similar approach to Grover \emph{et al.},
but we generalize the computation model to both parallel and serial
computation, and show how the energy of low block error rate decoders must
scale with block length $n$.  We believe that this approach can guide the
development of codes and decoding circuits that are optimal from an energy
standpoint.

In this paper, in Section \ref{sec:VLSI-Model} we will describe the VLSI
model that will be used to derive our bounds on decoding energy.  Our
results apply to a decoder for a standard binary erasure channel, which
will be formally defined in Section \ref{sec:Channel-Model}.  In Section
\ref{sec:Existing-Results} we will describe some terminology and some key
lemmas used in our paper. The main contribution of this paper will be
given in Section \ref{sec:Key-Result-of} where we describe a scaling rule
for codes with long block length that have asymptotic error probability
less than $\half$. The approach used in this section is extended in Section
\ref{sec:SerialComp} to find a scaling rule for serial computation. Then,
in Section \ref{generalCase} we extend the approaches of the previous
sections to derive a non-trivial super-linear lower bound on circuit energy
complexity for a series of decoders in which the number of output pins can
vary with increasing block length.  These results are applied to find a
scaling rule for the energy of capacity approaching decoders as a function
of fraction of capacity in Section \ref{Sec:Consequences}.  We then give a
simple example in Section \ref{sec:upperBound} showing how an LDPC decoder
can be implemented with at most $O\left(n^2\log \log n\right)$ energy,
providing an upper bound to complement our fundamental lower bound.

\emph{Notation:} To aid our discussion of scaling laws, we use standard
Big-Oh and Big-Omega asymptotic notation, which is well discussed in
\cite{knuth1976big}.  We say that a function
$f\left(x\right)=O\left(g\left(x\right)\right)$ if and only if there is
some $M$ and some $x_{0}$ such that for all $x\ge x_{0}$,
$f\left(x\right)\le Mg\left(x\right)$. Similarly, we say that
$f\left(x\right)=\Omega\left(g\left(x\right)\right)$ if and only if
$g\left(x\right)=O\left(f\left(x\right)\right)$. Intuitively, this means
that the function $f\left(x\right)$ grows at least as fast (in an order
sense) as $g\left(x\right)$ and hence we use it with lower bounds. In the
following, a sequence of values $b_{1},b_{2},\ldots,b_{k}$ is denoted
$b_{1}^{k}$. Random variables are denoted with upper case letters; values
in their sample spaces are denoted with lower-case letters.

\section{VLSI Model\label{sec:VLSI-Model}}

\subsection{Description of Model}

The VLSI model that we will use is based on the work of
Thompson~\cite{Thompson}, and was used by El Gamal in \cite{ElGamal} and
Grover \emph{et al.} in \cite{groverFundamental}. The model consists of a
basic set of axioms that describe what is meant by a VLSI circuit and a
computation.  The model then relates two parameters of the circuit and
computation, namely the number of clock cycles used for the computation and
the area of the circuit, to the energy used in the computation.  Thompson
used this model to compute fundamental bounds on the energy required to
compute the discrete Fourier transform, as well as other standard
computational problems, including sorting. The results in this paper apply
to any circuit implemented in a way that is consistent with these axioms,
listed as follows:

\begin{figure} 
\centering
\includegraphics{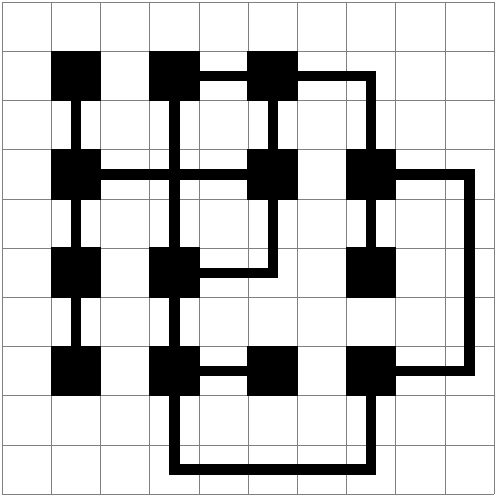}
\caption{Diagram of a possible VLSI circuit. The circuit is laid out on a
grid of squares and the squares of the grid that are filled in fully
represent computational nodes and the thinner lines represent wires. Also
present is a wire-crossing in one of the grid squares. The area of this
circuit is proportional to the number of grid-squares that contain either a
computational node, a wire, or a wire crossing. } \label{circuit}
\end{figure}

\begin{itemize}
\item A circuit consists of two types of components: wires and nodes. In
such a circuit, wires carry binary messages and nodes can perform simple
computations (\textit{e.g.}, \textsc{and}, \textsc{xor}), all laid out on a
grid of squares. Wires carry information from one node to another node.
Wires are assumed to be bi-directional (at least for the purpose of lower
bounds). In each clock cycle, each node sends one bit to each of the nodes
it is connected to over the wire. We in general allow a node to perform a
random function on its inputs. In a deterministic function, the output of
the function is determined fully by its inputs. By a random function we
mean that the outputs of a particular node, conditioned on the inputs being
some element from the set of possible inputs, is a distribution
$p_{Y|X}\left(\cdot|\cdot\right)$ where $Y$ is a random variable
representing the possible outputs of the node and $X$ a random variable
representing the inputs of a node. In the particular case of a node with
$4$ input wires (and thus $4$ output wires because of our bidirectional
assumption on the wires) the random variables $X$ and $Y$ can take on
values from $\left\{ 0,1\right\} ^{4}$.

\item A VLSI circuit is a set of computational nodes connected to each
other using finite-width bi-directional wires. At each clock cycle, nodes
communicate with the other nodes to which they are connected. Computation
terminates at a pre-determined number, $\tau$, of clock cycles.

\item Wires run along edges and cross at grid points. There is only one
wire per edge. Each grid point contains a logic element, a wire crossing, a
wire connection, or an input/output pin. Grid points can be empty. A
computational node is a grid point that is either a logic element, a
wire-connection, or an input/output pin.

\item The circuit is planar, and each node has at most $4$ wires leading
from it.

\item Inputs of computation are stored in source nodes and outputs are
stored in output nodes. The same node can be a source node and an output
node. Each input can enter the circuit only at the corresponding source
node.

\item Each wire has width $\lw$. Any two wires are separated by at least
the wire-width. Any grid points are separated by distance at least $\lw$.
Each node is assumed to require at least $\lw^{2}$ wire area (length and
width at least $\lw$). 

\item Processing is done in ``batches,'' in which a set of inputs is
processed and outputs released before the next set of inputs arrive.

\item Energy consumed in a computation is proportional to $\Ac\tau$ where
$\Ac$ is the area occupied by the wires and nodes of the circuit and $\tau$
is the number of clock cycles required to execute the computation.
Precisely, the energy is assumed to be $\frac{1}{2}\Cc \Vdd^{2}\tau$ where
$\Cc=\Cu \Ac$, where $\Cu$ is the capacitance per unit wired area of the
circuit, and $\Vdd$ is the voltage used to drive the circuit. The quantity
$\frac{1}{2}\Cu \Vdd^{2}$ is denoted by $\tech$, which is the ``energy
parameter'' of the circuit. Processing energy for computation, $\Eproc$, is
thus given by $\Eproc=\tech\Ac\tau$. Since energy of a computation in our
model and the area time complexity are essentially the same, in this paper
we use the terms ``energy complexity'' and ``Area-Time'' complexity
interchangeably.
\end{itemize}

\subsection{Discussion of Model\label{sub:Discussion-of-Model}}

The circuit model described above allows us to consider a circuit as a
graph in which the computational nodes correspond to the nodes of a graph
and the wires correspond to edges.

The last assumption of our model, which relates the area and number of
clock cycles to energy consumed in a circuit, assumes that a VLSI circuit
is fully charged and then discharged to ground during each clock cycle.
Since the wires in a circuit are made of conducting material and are laid
out essentially flat, the circuit will have a capacitance proportional to
the area of the wires. Assuming that all the wires will need to be charged
at each clock cycle, there must be $\frac{1}{2}\Cc \Vdd^{2}$ energy
supplied to the circuit. For now, we do not consider what will happen if at
each clock cycle the state of some of the wires does not change. In the
literature (see \cite{Howard} and \cite{Rabaey}) this model is often used
to understand power consumption in a digital circuit so we do not attempt
to alter these assumptions for the purposes of this paper. Sometimes
leakage current of the circuit is factored into such models; we also
neglect this as we assume the frequency of computation is high enough so
that the power used in computation dominates.

There has been some work to understand the tradeoff between computational
complexity and code performance. One such example is \cite{WeiYuBenSmith},
in which the complexity of a Gallager Decoding Algorithm B was optimized
subject to some coding parameters. This however does not correspond to the
energy of such algorithms.

In \cite{Thompson} it was proven that the Area-Time complexity of any
circuit implemented according to this VLSI model that computes a Discrete
Fourier Transform must scale as $\Omega\left(n^{1.5}\log n\right)$.
However, there exist algorithms that compute in $O\left(n\log n\right)$
operations (for example, see \cite{CooleyTukey}); Thompson's results thus
imply that, for at least some algorithms, energy consumption is \emph{not}
merely proportional to the computational complexity of an algorithm. 

In the field of coding theory, Grover \emph{et al.}  in
\cite{groverFundamental} provided an example of two algorithms with the
same computational complexity but different computational energies. The
authors looked at the girth of the Tanner graph of an LDPC code. The girth
is defined as the minimum length cycle in the Tanner graph that represents
the code. They showed, using a concrete example, that for (3, 4)-regular
LDPC codes of girth 6 and 8 decoded using the Gallager-A decoding
algorithm, the decoders for girth 8 codes can consume up to 36\% more
energy than those for girth 6 codes. The girth of a code does not
necessarily make the decoding algorithm require more computations
(\textit{i.e.}, it does not increase computational complexity), but, for
this example, it \emph{does} increase the energy complexity.  This is
because codes with greater girth require the interconnection between nodes
to be more complex, even though the same number of computational nodes and
clock cycles may be required. This drives up the area required to make
these interconnections, and thus drives up the energy requirements. Also in
the field of coding theory, the work of Thorpe \cite{1023755} has shown
that a measure of wiring complexity of an LDPC decoder can be traded off
with decoding performance.

 Thus, current research  suggests that in decoding algorithms there appears
to be a  fundamental trade-off between code performance and interconnect
complexity. This paper attempts to find an analytical characterization of
this trade-off.

Our paper considers a generic model of computation, but of course it does
not completely reflect all methods of implementing a circuit. We discuss
some circuit design techniques that our model does not directly consider
below.

\subsubsection{Multiple Layer VLSI Circuits}
Modern VLSI circuits differ from the Thompson model in that the number of
VLSI layers is not one (or two if one counts a wire crossing as another
layer). Modern VLSI circuits allow multiple layers. Fortunately, it is
known that if $L$ layers are allowed, then this can decrease the total area
by at most a factor of $L^{2}$ (see, for example, \cite{ThompsonThesis} or
\cite{876069}). For the purposes of our lower bounds, if the number of
layers remains constant as $n$ increases, we can modify our energy lower
bound results by dividing the lower bounds by $L^{2}$.  If, however, the
number of layers can grow with $n$ our results may no longer hold. Note
also that this only holds for the purpose of lower bound. It may not be
possible to implement a circuit with an area that decreases by a factor of
$L^{2}$, and so the upper bounds of Section \ref{sec:upperBound} cannot be
similarly modified.

\subsubsection{Adiabatic Computing}
The model used in this paper assumes that after every clock cycle the
circuit is entirely discharged and the energy used to charge the circuit is
lost. There exists extensive research into circuit designs in which this is
not the case (for an overview of this type of computing, called adiabatic
computing, see \cite{Denker}). Our results do not apply to such circuit
designs.

\subsubsection{Using Memory Elements in Circuit Computation}
The Thompson model does not allow for the use of special memory nodes in
computation that can hold information and compute the special function of
loading and unloading from memory. Such a circuit can be created using the
Thompson model, but it may be that a strategic use of a lower energy memory
element can decrease the total energy of a computation. Intuitively,
however, the use of a memory element to communicate information within a
circuit should still be proportional to the distance that information is
communicated. Grover in \cite{GroverInfoFriction} proposed a ``bit-meters''
model of energy computation and derives scaling rules similar to our
results, suggesting that, at least in an order sense, the circuit model we
use is general enough to understand the scaling of high block length codes
even if lower energy memory is used. Understanding precisely what kind of
gain the use of a memory element can provide in energy complexity is beyond
the scope of this paper.

\section{Channel Model\label{sec:Channel-Model}}

We will consider a noisy channel model that is similar to the model used by
Grover \emph{et al.} in \cite{groverFundamental}. An information sequence
of $k$ independent fair coin flips $b_{1}^{k}\in\left\{ 0,1\right\} ^{k}$
is encoded into $2^{nR}$ binary-alphabet codewords $x_{1}^{n}\in\left\{
0,1\right\} ^{n}$; and thus this code has rate $R=k/n$ bits/channel use.
The sequence $x_{1}^{n}$ is passed through a binary erasure channel with
erasure probability $\epsilon$, resulting in a received vector
$y_{1}^{n}\in\left\{ 0,1,?\right\}^n $, where the $?$ symbol corresponds to
the erasure event.

 The decoder estimates the input sequence $\hat{b}_{1}^{k}\in\left\{
0,1\right\} ^{k}$ by computing a function of the received vector
$y_{1}^{n}$. The outputs of the noisy channel thus become inputs into the
decoder circuit. In our most general model of computation, it is required
that these channel outputs are eventually input into an input node of the
circuit. In a parallel implementation model used for Theorem
\ref{mainTheorem}, each of these $n$ decoder input symbols are input, at
the beginning of the computation, into the $n$ input nodes of the decoder.
In a more general computational model used in Theorems \ref{serialTheorem}
and \ref{generalTheorem}, it is assumed that each of these symbols are
input into the input nodes of the decoder during some clock cycle in the
computation. Note that we allow each of the $n$ symbols to be inserted into
the circuit at any input node location during any clock cycle, but we also
require, according to our model, that each input is injected only once into
the circuit. Thus, our model does not subsume circuit implementations that,
at no cost, allow the same input symbol to be inserted into the circuit in
multiple places. 
 
The probability of block error is defined as
\[
P_{e}^{\text{blk}}=\Pr\left(\hat{b}_{1}^{k}\neq b_{1}^{k}\right).
\]
The lower bounds used in our result are valid for a binary erasure channel,
but also for any channel that can result from a degraded version of a
binary erasure channel.  Hence, if we let $\epsilon=2\pch$ then our results
apply to lower bounds on decoders for binary symmetric channels with
crossover probability $\pch$.

\section{Definitions and Main Lemmas\label{sec:Existing-Results}}

The main theorems in this paper rely on the evaluation of a particular
limit, which we present as a lemma below.
\begin{lem} \label{limitLemma}
For any constant $c$, $0<c<1$, and any constant $c'>0$:
\begin{equation}\label{eq:keyLimitUsed}
\lim_{n\rightarrow\infty}\left(1-\exp\left(-c\log n\right)\right)^{\frac{c'n}{\log n}}=0.
\end{equation}
\end{lem}
\begin{IEEEproof}
This result follows simply from taking the logarithm of the expression in
(\ref{eq:keyLimitUsed}) and using L'H\^opital's rule to show that the
logarithm approaches $-\infty$.
\end{IEEEproof}

\subsection{Relation Between Energy and Bits Communicated\label{sub:Relation-Between-Energy}}

Grover \emph{et al.} in \cite{groverFundamental} used a nested bisection
technique, involving subdividing circuits, to derive the two key lemmas
used in this paper.  A circuit created according to the Thompson model can
be considered a graph in which the computational nodes correspond to graph
vertices and the wires are graph edges.  To understand these lemmas we must
first understand what is meant by (a) a minimum bisection of a circuit and
(b) a nested bisection of a circuit.  Informally, a bisection of a graph is
a set of edges whose removal results in at least two graphs of essentially
the same size that are unconnected to each other. A bisection can also be
defined in terms of separating a particular subset of vertices. A formal
definition is given below.

\begin{figure} 
\centering
\raisebox{3ex}{\includegraphics{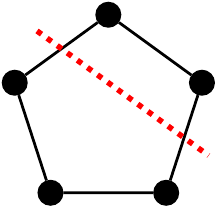}}~\includegraphics{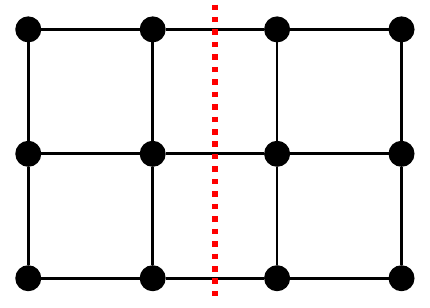}
\caption{Example of two graphs with a minimum bisection labelled. Nodes are
represented by circles and edges by lines joining the circles. A dotted
line crosses the edges of each graph that form a minimum bisection.}
\label{minimumBisectionWidth}
\end{figure}
\begin{defn}
Let $G=\left(V,E\right)$ be a graph. Let $S\subseteq V$ be a subset of the
vertices, and $E_{s}\subseteq E$ be a subset of the edges. Then $E_{s}$
\textsl{bisects} $S$ in $G$ if deletion of the edges in $E_{s}$ from the
graph cuts $V$ into disconnected sets $V_{1}$ and $V_{2}$, with $V_{1}\cup
V_{2}=V$,  and $S$ into sets $S_{1}\subseteq V_{1}$ and $S_{2}\subseteq
V_{2}$ such that $\left|\left|S_{1}\right|-\left|S_{2}\right|\right|\le1$.
A bisection of $V$ in $G$ is called a bisection of $G$.

A minimal bisection is a bisection of a graph whose size is minimal over
all bisections. The \textsl{minimum bisection width} is the size of a
minimal bisection. For a general graph, finding a minimum bisection is a
difficult problem (in fact, it is NP-Complete \cite{Garey1976237}), but all
that is required for the results we will use is that a minimum bisection
exists for every graph.  Fig.~\ref{minimumBisectionWidth} shows minimal
bisections of a few simple graphs.
\end{defn}

Note that the definition of a minimum bisection also applies to subsets of
the vertices of a graph.  A circuit has both input nodes and output nodes.
The input (resp. output) nodes of the graph corresponding to a circuit are
those nodes of the graph that correspond to the input (resp. output) nodes
of the circuit.  For the purposes of the results in this paper, we will
consider bisections of the graph that bisect the output nodes. 

We will also be using some other terms throughout the paper which we define
below.

\begin{figure} 
\centering
\includegraphics{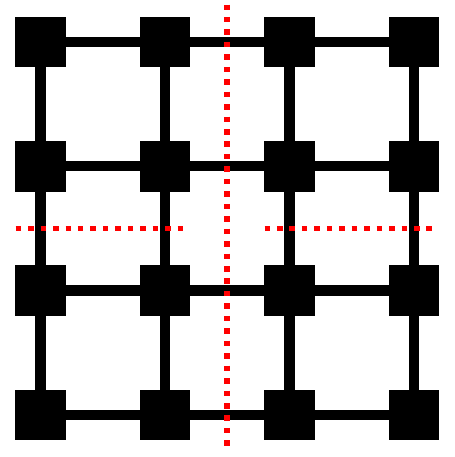}

\caption{Example of a possible circuit undergoing two stages of nested
minimum bisections. The dotted line down the middle is a first nested
bisection, and the other two horizontal dotted lines are the bisections
that divide the two subcircuits that resulted from the first stage of the
nested bisections, resulting in four subcircuits. We are concerned with the
number of bits communicated across $r$-stages of nested minimum bisections.
In these two stages of nested minimum bisections, we see that $8$ wires are
cut. Because we assume wires are bidirectional, and thus two bits are
communicated across these wires every clock cycle, in the case of this
circuit we have $B_r=8\times2\times\tau$, where $\tau$ is the number of
clock cycles. It will not be important how to actually do these nested
bisections, rather it is important only to know that any circuit can
undergo these nested bisections. } \label{nestedBisectionExamples}
\end{figure}

If one performs a minimum bisection on the output nodes of the graph
corresponding to the interconnection graph of a circuit, this results in
$2$ disconnected subcircuits. The output nodes of these two (now
disconnected) subcircuits can thus each be minimally-bisected again,
resulting in $4$ subcircuits. 
\begin{defn}
This process, when repeated $r$ times, is said to perform an
\textsl{$r$-stage nested minimum bisection}.  Note that this divides the
graph into $2^{r}$ disconnected components, which we will refer to as
\textsl{subcircuits}. 
\end{defn}
When a circuit is viewed as a graph, a subcircuit can be viewed as a
subgraph induced by a nested minimum bisection. When viewed as a circuit
according to the Thompson model, it is a collection of computational nodes
joined by wires laid out on a grid pattern.  An example of a mesh-like
circuit with 16 nodes undergoing two stages of nested minimum bisections is
shown in Fig.~\ref{nestedBisectionExamples}.

After a circuit undergoes $r$-stages of nested minimum bisections,  label
each of the $2^{r}$ subcircuits with a unique integer $i$ in which $1\le
i\le2^{r}$. Consider a particular subcircuit $i$. During the $r$-stages of
nested minimum bisections, a number of edges $f_{i}$ are removed that are
incident on the graph corresponding to subcircuit $i$ (we can think of this
as the ``fan-out'' of this subcircuit).

\begin{defn} \label{def:biDefn}
During the course of the computation, the number of bits communicated to
subcircuit $i$ is $b_i=\tau f_{i}$, where $\tau$ is the number of clock
cycles, and we refer to $b_i$ as the \textsl{bits communicated to
subcircuit $i$ during a computation}. 
\end{defn}
The quantity $b_i$ is associated with a particular subcircuit induced by a
particular $r$-stage nested minimum bisection. When discussed, this
quantity's association with a particular $r$-stage nested minimum bisection
is implicit.

\begin{defn}\label{defn:Br}
The quantity $B_{r}=\sum_{i=1}^{2^{r}}b_{i}$ denotes the  \textsl{number
of bits communicated across all edges deleted in an $r$-stage nested
minimum bisection.} Note that this is a quantity associated with a
particular $r$-stage nested minimum bisection.
\end{defn}

The quantity $B_{r}$ will be important in the proofs of the theorems in the
paper. Specifically, it can be shown that if a decoding circuit (which we
will define below) has a large $B_{r}$ for a particular $r$-stage nested
minimum bisection, then the energy expended during a computation by this
circuit must be high. As well, it can be shown that if this quantity is
low, then the probability that the circuit makes an error is high.

The above definitions are general and can apply to the computation of any
function. However, for our bounds we will be finding lower bounds on the
energy complexity of decoding circuits.

\begin{defn} An $\left(n,k\right)$ \textsl{parallel decoding circuit} is a
circuit that has $n$ input nodes (accepting symbols in $\left\{
0,1,?\right\} $) and $k$ output nodes (producing symbols in $\left\{
0,1\right\} $). The $n$ input nodes are to receive the outputs of a noisy
channel (which for the purposes of lower bound we assume to be a binary
erasure channel with erasure probability $\epsilon$). At the end of the
computation the decoder is to output the estimate of the original codeword.
Note that this circuit decodes a rate $R=k/n$ code.  \end{defn} Note that
in the Thompson model it is assumed that all inputs are binary. For the
purposes of lower bound, we allow for the inputs into the computation to be
either $0$, $1$ or $?$, where $?$ is the erasure symbol. At every clock
cycle, we allow input nodes to perform a function on their input symbol, as
well as the bits input into the node at the clock cycle. These nodes may
then output any function of these inputs along the wires leading from the
node.

This definition will be generalized to serial computation models in the
discussions preceding Theorem~\ref{serialTheorem}. Note also that our model
of a decoding circuit allows for an input to be an erasure symbol, which is
a slight relaxation of the Thompson circuit model. However, in our
theorems, the key point will be that, if in a particular subcircuit all the
$n_i$ input nodes of that subcircuit are erased, then, \textsl{conditioned
on this event}, the distribution of the possible original inputs to the
channel of the $k_i$ bits that a subcircuit $i$ is to estimate remains
uniform. This is a result of the symmetric nature of a binary erasure
channel, and this will allow us to directly apply
Lemma~\ref{pidgeonholeLemma} to form lower bounds on probability of error.

After a decoding circuit undergoes $r$-stages of nested minimum bisections,
each subcircuit will have roughly an equal number of output nodes, but the
number of input nodes may vary (the actual number of input nodes in each
subcircuit in general will be a function of the particular graph structure
of the circuit, and the particular $r$-level nested minimum bisection
performed).  \begin{defn}\label{def:niDefn} We refer to this quantity as
the \textsl{number of input nodes in subcircuit $i$} and denote it $n_{i}$.
\end{defn} Note that this quantity is determined by the particular
$r$-stage nested minimum bisection, but for notational convenience we will
consider the relation of this quantity to the particular structure of the
$r$-stage nested minimum bisection to be implicit.

\begin{defn} A particular $i$th subcircuit formed by an $r$-stage nested
minimum bisection will have a certain number of output nodes, which we
denote $k_i$. This quantity is referred to as the \textsl{number of output
nodes in subcircuit $i$}.  \end{defn} In a fully parallel computation model
(which we will employ in Theorem~\ref{mainTheorem}), at the end of the
computation, these output nodes are to hold a vector $\hat{k_{i}}\in
\left\{ 0,1\right\} ^{k_{i}}$, where the vector to be estimated is a vector
$k_{i}\in  \left\{ 0,1\right\} ^{k_{i}}$ which is produced by a series of
fair coin flips as described in Section~\ref{sec:Channel-Model}. Since at
the end of the computation these output nodes are to hold an estimate of a
vector of length $k_i$  it is said that in this case $k_i$ is the
\textsl{number of bits responsible for decoding by subcircuit $i$}. The
probability of error for a subcircuit is precisely the probability that,
after the end of the computation, $\hat{K}_{i}\ne K_{i}$.

\begin{lem} \label{pidgeonholeLemma}
Suppose that $X$, $Y$, and $\hat{X}$ are random variables that form a
Markov chain $X\rightarrow Y\rightarrow\hat{X}$. Suppose furthermore that
$X$ takes on values from a finite alphabet $\mathcal{X}$ with a uniform
distribution (i.e., $P\left(X=x\right)=\frac{1}{\left|\mathcal{X}\right|}$
for any $x\in\mathcal{X}$) and $Y$ takes on values from an alphabet
$\mathcal{Y}$. Suppose furthermore that $\hat{X}$ takes on values from a
set $\mathcal{\hat{X}}$ such that $\mathcal{X}\subseteq\mathcal{\hat{X}}$.
Then,
\[
P\left(\hat{X}=X\right)\le\frac{\left|\mathcal{Y}\right|}{\mathcal{\left|X\right|}}.
\]
\end{lem}
\begin{rem}
This general lemma is meant to make rigorous a fundamental notion that will
be used in this paper. As applied to our decoding problem, the random
variable $X$ can be thought of as the input to a binary erasure channel,
and $Y$ can be any inputs into a subcircuit of a computation, and $\hat{X}$
can be thought of as a subcircuit's estimate of $X$. This lemma makes
rigorous the notion that if a subcircuit has fewer bits input into it than
it is responsible for decoding, then the decoder must guess at least $1$
bit, and makes an error with probability at least $\half$.  This scenario
is actually a special case of this lemma in which
$\left|\mathcal{Y}\right|=2^{m}$ and $\left|\mathcal{X}\right|=2^{k}$ for
integers $k$ and $m$, where $m<k$.
\end{rem}
\begin{IEEEproof}
Clearly, by the law of total probability,

\begin{align*}
P\left(X=\hat{X}\right) & =  \sum_{x\in\mathcal{X}}\sum_{y\in\mathcal{Y}}P_{X,Y,\hat{X}}\left(x,y,x\right)\\
 & =  \sum_{x\in\mathcal{X}}\sum_{y\in\mathcal{Y}}P_{X}\left(x\right)P_{Y|X}\left(y|x\right)P_{\hat{X}|Y}\left(x|y\right)
\end{align*}
where we simply expand the term in the summation according to the
definition of a Markov chain. Using
$P_{X}\left(x\right)=\frac{1}{\left|\mathcal{X}\right|}$ we get:
\[
P\left(X=\hat{X}\right)=\frac{1}{\left|\mathcal{X}\right|}\sum_{x\in\mathcal{X}}\sum_{y\in\mathcal{Y}}P_{Y|X}\left(y|x\right)P_{\hat{X}|Y}\left(x|y\right)
\]
and using $P_{Y|X}\left(y|x\right)<1$ because it is a probability, and
changing the order of summation gives us:
\[
P\left(X=\hat{X}\right)\le\frac{1}{\left|\mathcal{X}\right|}\sum_{y\in\mathcal{Y}}\sum_{x\in\mathcal{X}}P_{\hat{X}|Y}\left(x|y\right).
\]
Since $\sum_{x\in\mathcal{X}}P_{\hat{X}|Y}\left(x|y\right)\le1$ (as we are
summing over a subset of values that $\hat{X}$ can take on), we get:
\[
P\left(X=\hat{X}\right)\le\frac{1}{\left|\mathcal{X}\right|}\sum_{y\in\mathcal{Y}}1=\frac{\left|\mathcal{Y}\right|}{\mathcal{\left|X\right|}}.
\]
\end{IEEEproof}

In the proofs of the theorems in this paper, we will be dividing a circuit
up into pieces and then we will let $n$ grow larger. Technically, a circuit
can only be divided into an integer fraction of pieces.  However, in most
cases this does not matter. To make this notion rigorous, we will need to
use the following lemma:

\begin{lem}\label{roundingErrorLemma} 
Let $h:\mathbb{R}\rightarrow\mathbb{R}$ be a function such that
$\left|h\left(x\right)-x\right|\le a$ for sufficiently large $x$ and some
positive constant $a$. If there are functions
$f,g:\mathbb{R}\rightarrow\mathbb{R}$, and $g$ is continuous for
sufficiently large $x$, and if
$\lim_{x\rightarrow\infty}f\left(g\left(x\right)\right)=c$ for some
constant $c\in \mathbb{R}$, and if $\lim_{x\rightarrow\infty}g\left(x\right)=\infty$
then
$\lim_{x\rightarrow\infty}f\left(h\left(g\left(x\right)\right)\right)=c$.
\end{lem}

\begin{IEEEproof} 
Suppose 
\[
\lim_{x\rightarrow\infty}f\left(g\left(x\right)\right)=c.
\]

To show that
$\lim_{x\rightarrow\infty}f\left(h\left(g\left(x\right)\right)\right)=c$ we
need to construct, given some $\epsilon$, a particular $x_{0}$ such that
for all $x>x_{0},$
$\left|f\left(h\left(g\left(x\right)\right)\right)-c\right|<\epsilon$.
Since $g$ grows unbounded, and is continuous for sufficiently large $x$,
then there must be a particular value of $x$ (call it $x'$) such that
$g\left(x\right)$ takes on all values greater than $g\left(x'\right)$ for
some $x>x'$. As well, for any $\epsilon>0$ there exists some $x''$ such
that for all $x>x''$,
$\left|f\left(g\left(x\right)\right)-c\right|<\epsilon$.  In particular
this is true for some $x''>x'$. Thus, choose $x_{0}$ to be the least number
greater than $x''$ in which $g\left(x_{0}\right)=g\left(x''\right)+a$ (this
must exist because $g$ takes on all values greater than
$g\left(x''\right)$).  Thus, for $x>x_{0}$ $h\left(g\left(x\right)\right)$
only takes on values greater than $g\left(x''\right)$ (because
$\left|h\left(x\right)-x\right|\le a$).  Since
$\left|f\left(g\left(x\right)\right)-c\right|<\epsilon$ for all $x>x''$,
thus $\left|f\left(h\left(g\left(x\right)\right)\right)-c\right|<\epsilon$
for all $x>x_{0}$, since $h\left(g\left(x\right)\right)$ can only take on
values that $g\left(x\right)$ takes on for $x>x''$.
\end{IEEEproof}
\begin{cor}
This result applies when $h\left(\cdot\right)$ is the floor function,
denoted$\left\lfloor \cdot\right\rfloor $, since $\left|\left\lfloor
x\right\rfloor -x\right|\le1$.
\end{cor}

We will need to make one observation that will be used in the three main theorems
of this paper, which we present in the lemma below.
\begin{lem} \label{ConvexOptimizationLemma}
	If $\epsilon>0$ and $n_{1},n_{2},\ldots,n_{m}$ are positive integers
	subject to the restriction that $\sum_{i=1}^{m}n_{i}\le n$ then: 
	\[
	\prod_{i=1}^{m}\left(1-\epsilon^{n_{i}}\right)\le\left(1-\epsilon^{\frac{n}{m}}\right)^{m}
	\]
\end{lem}
\begin{IEEEproof}
	The proof follows from a simple convex optimization argument.
\end{IEEEproof}

Grover \emph{et al.} in \cite{groverFundamental} uses a nested bisection
technique to prove a relation between energy consumed in a circuit
computation and bits communicated across the $r$-stages of nested
bisections which we present as a series of two lemmas, the second which we
will use directly in our results.

\begin{lem} \label{AtausquaredLemma}
For a circuit undergoing $r$-stages of nested bisections, in which the
total number of bits communicated across all $r$-stages of nested
bisections is $B_{r}$, then 
\[
A\tau^{2}\ge\frac{\left(\sqrt{2}-1\right)^{2}}{16}\frac{B_{r}^{2}}{2^{r+1}}\lambda^{2} \label{equationWithAtau}
\]
where $A$ is the area of the circuit and $\tau$ is the number of clock
cycles during the computation.
\end{lem}
\begin{IEEEproof}
See \cite{groverFundamental} for a detailed proof. Here we provide a
sketch. To accomplish this proof, $r$-stages of nested minimum bisections on a circuit are performed
 and then a
principle due to Thompson \cite{Thompson} is applied that states that the
area of a circuit is at least proportional to the square of the minimum
bisection width of the circuit. Also, the number of bits communicated to a
subcircuit cannot exceed the number of wires entering that subcircuit
multiplied by the number of clock cycles. The area of the circuit (related
to the size of the minimum bisections performed) and the number of clock
cycles (more clock cycles allow more bits communicated across cuts) are
then related to the number of bits communicated across all the edges
deleted during the $r$-stages of nested bisections.
\end{IEEEproof}

\begin{lem}
\label{lem:EdecLemma} If a circuit as described in
Lemma~\ref{AtausquaredLemma} in addition has at least $\beta$ nodes, then
the $A\tau$ complexity of such a computation is lower bounded by:
\[
A\tau\ge\frac{\sqrt{2}-1}{4\sqrt{2}}\sqrt{\frac{\beta}{2^{r}}}B_{r}\lw^{2}
\]
\end{lem}
\begin{IEEEproof}
Following the same arguments of Grover \emph{et al.} in
\cite{groverFundamental} (which we reproduce to get the more general result
we will need), note that if there are at least $\beta$ computational nodes,
then
\[
A\ge\beta\lw^{2}
\]
which, when combined with Lemma~\ref{AtausquaredLemma} results in:
\[
A^{2}\tau^{2}  \ge \frac{\left(\sqrt{2}-1\right)^{2}}{16}\frac{B_{r}^{2}}{2^{r+1}}\beta\lw^{4}
\]
which yields the statement of the Lemma upon taking the square root.
\end{IEEEproof}

\begin{rem}
In terms of our energy notation, the result of Lemma~\ref{lem:EdecLemma}
implies that for such a circuit with at least $\beta$ computational nodes,
the energy complexity is lower bounded by:
\[
\Eproc\ge\tech\lw^{2}\frac{\sqrt{2}-1}{4\sqrt{2}}\sqrt{\frac{\beta}{2^{r}}}B_{r}=\Ktech\sqrt{\frac{\beta}{2^{r}}}B_{r}
\]
where $\Ktech=\tech\lw^{2}\frac{\sqrt{2}-1}{4\sqrt{2}}$.\label{BondForATau}
\end{rem}

\subsection{Bound on Block Error Probability\label{sub:Bound-on-Block}}

The key lemma that will be used in the first theorem of this paper is due
to Grover \emph{et al.}  \cite{groverFundamental}. We modify the lemma
slightly.

\begin{lem}
\label{lem:PEblockLemma} For any code implemented using the VLSI model for
an erasure channel with erasure probability $\epsilon$, for any
$r<\log_{2}\left(\frac{k}{2}\right)$,
\[
\text{either }P_{e}^{blk}\ge\frac{1}{2}-\left(1-\epsilon^{\frac{n_{i}}{2^{r-1}}}\right)^{2^{r-1}}\text{ or }B_{r}\ge\frac{k}{2}.
\]
\end{lem}

The proof uses the same approach as Grover \emph{et al. }in
\cite{groverFundamental} but we modify it slightly to ease the use of our
lemma for our theorem and to conveniently deal with the possibility that a
decoder can guess an output of a computation.

Let $s_i$ be the number of input bits erased in the $i$th subcircuit after
$r$-stages of nested bisections. Furthermore, recall from Definition
\ref{def:biDefn} that $b_{i}$ is the number of bits injected into the $i$th
subcircuit during the computation. Also, recall from Definition
\ref{def:niDefn} that $n_i$ is the number of input nodes located within the
$i$th subcircuit. We use the principle that if
\[
\frac{k}{2^{r}}<n_{i}-s_i+b_{i}
\]
for any subcircuit then the probability of block error is at least $\half$.
This is a very intuitive idea; if the number of bits that are not erased,
plus the number of bits injected into a circuit is less than the number of
bits the circuit is responsible for decoding, the circuit must at least
guess $1$ bit. This argument will be made formal in the proof that follows.

\begin{IEEEproof} (of Lemma \ref{lem:PEblockLemma})
Suppose that all the $n_i$ input bits injected into the $i$th subcircuit
are the erasure symbol. Then, conditioned on this event, the distribution
of the $k_i$ bits that this subcircuit is to estimate is uniform (owing to
the symmetric nature of the binary erasure channel). Furthermore, if
$b_{i}<\frac{k}{2^{r}}$ then the number of bits injected into the
subcircuit is less than the number of bits the subcircuit is responsible
for decoding. Combining these two facts allows us to apply
Lemma~\ref{pidgeonholeLemma} directly to conclude that, in the event all
the inputs bits of a subcircuit are erased, and the number of bits injected
into the subcircuit is less than $\frac{k}{2^{r}}$, then the subcircuit
makes an error with probability at least $\half$. Denote the event that all
inputs bits in subcircuit $i$ are erased as $W_{i}^{r}$. The probability of
this event is given by
\[
P\left(W_{i}^{r}\right)=\epsilon^{n_{i}}.
\]

Suppose that $B_{r}<k/2$ (where we recall from Definition \ref{defn:Br}
that $B_{r}$ is the total number of bits communicated across all edges cut
in $r$-stages of nested minimum bisections).  Let $S=\left\{
i:b_{i}<\frac{k}{2^{r}}\right\} $ be the set of indices $i$ in which
$b_{i}$ (the bits communicated to the $i$th subcircuit) is smaller than
$\frac{k}{2^{r}}$. We first claim that $\left|S\right|>2^{r-1}$.  To prove
this claim, let $\bar{S}=\left\{ i:b_{i}\ge\frac{k}{2^{r}}\right\} $ and
note that $\frac{k}{2}>B_{r}=\sum
b_{i}\ge\sum_{i\in\bar{S}}\frac{k}{2^{r}}=\left|\bar{S}\right|\frac{k}{2^{r}}$,
from which it follows that $\left|\bar{S}\right|<2^{r-1}$. Since
$\left|S\right|+\left|\bar{S}\right|=2^{r}$, the claim follows. 

Hence, in the case that $B_{r}\le k/2$, because of the law of total
probability:
\begin{align}
P\left(\text{correct}\right) &=  P\left(\cap_{i\in S}\bar{W_{i}^{r}}\right)P\left(\text{correct}|\cap_{i\in S}\bar{W_{i}^{r}}\right)+P\left(\cup_{i\in S}W_{i}^{r}\right)P\left(\text{correct}|\cup_{i\in S}W_{i}^{r}\right)\nonumber \\
 & \le  \prod_{i\in S}\left(1-\epsilon^{n_{i}}\right)+\frac{1}{2}\label{eq:usingLawOfTotal}
\end{align}
where the event $\cap_{i\in S}\bar{W_{i}^{r}}$ is the event that each of
the subcircuits indexed in $S$, after $r$-stages of nested bisections, do
not have all their $n_{i}$ input bits erased. We then note that, in this
case, the probability of the circuit being decoded correctly is at most
$1$. For the second term, we note that conditioned on the event that at
least one of the subcircuits indexed in $S$ has all their input bits
erased, since the circuit must at least guess $1$ bit, the probability of
the circuit decoding successfully is at most $\half$, by Lemma
\ref{pidgeonholeLemma}.

Since $\sum_{i\in S}n_{i}\le\sum_{i=1}^{2^{r}}n_{i}=n$, subject to this
restriction, Lemma \ref{ConvexOptimizationLemma} shows the expression in
(\ref{eq:usingLawOfTotal}) is maximized when
$n_{i}=\frac{n}{\left|S\right|}$ for each subcircuit in $S$.  Hence,
\[
P\left(\text{correct}\right)\le\left(1-\epsilon^{\frac{n}{\left|S\right|}}\right)^{\left|S\right|}+\frac{1}{2}.
\]
Thus, either $B_{r}\le\frac{k}{2}$ or $\left|S\right|\ge2^{r-1}$ which
implies
\[
P\left(\text{correct}\right)\le\left(1-\epsilon^{\frac{n}{2^{r-1}}}\right)^{2^{r-1}}+\frac{1}{2}
\]
and so
\begin{align*}
P_{e}^{blk} & =  1-P\left(\text{correct}\right)\\
 & \ge  \frac{1}{2}-\left(1-\epsilon^{\frac{n}{2^{r-1}}}\right)^{2^{r-1}}.
\end{align*}
\end{IEEEproof}

\section{Key Result: A Fundamental Scaling Rule for Energy of Low Block Error Probability Decoders\label{sec:Key-Result-of}}

We define a coding scheme as a sequence of codes of fixed rate
 together with decoding circuits, in which the block length of each
successive code increases. 
We define $P_{e}^{\text{blk}, n}$ as the block error
probability for the decoder of block length $n$ in this scheme. An example
of a coding scheme would be a series of regular LDPC codes together
with LDPC decoding circuits in which the block length $n$ doubles for each
decoder in the sequence.
Our results are general and would apply to any particular
coding scheme using any circuit implementation and any decoding algorithm.
The key result of this paper is given in the following theorem:
\begin{thm} \label{mainTheorem}
For every coding scheme in which
$\lim_{n\rightarrow\infty}P_{e}^{\text{blk}, n}<\half$,
there exists some $n_0$ such that for all $n>n_0$, for any circuit implemented
according to the VLSI model with parameters $\tech$ and $\lw$,
\begin{equation}
\Edec>\Ktech\sqrt{\frac{\left(\log_{2}n\right)}{\log_{2}\left(\frac{1}{\epsilon}\right)}}\frac{Rn}{2}\label{eq:TheoremToProve}
\end{equation}
where $\Edec$ is the energy used in the decoding and $\Ktech=\tech\lw^{2}\frac{\left(\sqrt{2}-1\right)}{4\sqrt{2}}$ is a constant that
depends on circuit technology parameters that we defined before.\end{thm}
\begin{rem}
The requirement that $\lim_{n\rightarrow\infty}P_{e}^{\text{blk}, n}<\half$ for
our bound in (\ref{eq:TheoremToProve}) though reasonable, is not necessary for a good design.
Typical block error probability requirements
may be on the order of $10^{-5}$ or $10^{-6}$, although if the block error probabilities are
lower bounded by $\half$ for a series of decoding schemes,
this is not necessarily a bad design. The individual bit error probabilities
(the probability that a randomly selected output bit of the decoder
is decoded correctly) may indeed be acceptably low. However, our results
do not consider such schemes. It is also not necessary for a decoding scheme
to have a block length that gets arbitrarily large. However, a capacity-approaching
code must have block length that approaches infinity and our result can be used to understand how the
energy complexity of such decoding algorithms approach infinity.\end{rem}
\begin{IEEEproof}
The theorem follows from an appropriate choice for $r$, the number
of nested bisections we perform. We can choose any nonnegative integer
$r$ so that $r<\log_{2}\left(\frac{k}{2}\right)$. Note that $k=nR$
is the number of bits the decoder is responsible for decoding. As
$k$ gets large, we can thus choose any $r$ so that $1\le2^{r}\le\frac{k}{2}$.
Thus, we choose an $r$ so that, approximately, $2^{r}=\frac{2\log\left(\frac{1}{\epsilon}\right)n}{K\log n}$, for a 
value of $K$ which we will choose later.
In particular, we will choose $r=\left\lfloor \log_{2}\left(\frac{2\log\left(\frac{1}{\epsilon}\right)n}{K\log n}\right)\right\rfloor $. 

This is valid so long as $n$ is sufficiently large, for some $0<K<1$.
 Note that $\log\frac{1}{\epsilon}>0$
since $0<\epsilon<1$. Since $\frac{k}{2}=\frac{Rn}{2}$ , this is
a valid choice for $r$ so long as
\[
\frac{2\log\left(\frac{1}{\epsilon}\right)n}{K\log n}<\frac{R}{2}
\]
which must occur as the left side of the inequality approaches $0$
as $n$ gets large. We can plug this value for $r$ into Lemma~\ref{lem:PEblockLemma},
but we will simplify the expression by neglecting the floor function, as application of Lemma~\ref{pidgeonholeLemma}
will show that this does not alter the evaluation of the limit that
we will compute, as we can see our choice for $r$ grows unbounded
with $n$. Thus, either
\begin{equation}
P_{e}^{blk,n}\ge\frac{1}{2}-\left(1-\exp\left(-K\log n\right)\right)^{\frac{1}{K}\frac{\log \left(\frac{1}{\epsilon}\right)n}{\log n}}
\label{eq:BoundToFindLimitOf}
\end{equation}
or, applying Lemma~\ref{lem:EdecLemma} by recognizing that there are at least $\beta=n$ nodes, 
\begin{equation}
\Edec>\Ktech\sqrt{\frac{\left(K\log n\right)}{\log \left(\frac{1}{\epsilon}\right)}}\frac{Rn}{2}. \label{IfLimitIsNotOverHalf}
\end{equation}
By a direct application of Lemma~\ref{limitLemma},  so long as $K<1$, the bound in (\ref{eq:BoundToFindLimitOf}) approaches $\half$ which we can see as follows:
\begin{align*}
\lim_{n\rightarrow\infty}P_{e}^{\text{blk}, n}
&\ge \frac{1}{2}-\lim_{n\rightarrow\infty}\left(1-
\exp\left(-K\left(\log n\right)\right)
\right)^{\frac{ Kn}{\left(\log n\right)}}\\
&=\frac{1}{2}.
\end{align*}
This implies
that, in the limit of large block sizes, the probability of block error must be lower bounded by $\half$, unless $B_r\ge\frac{k}{2}$.
But then by (\ref{IfLimitIsNotOverHalf}), it must be that 
\begin{equation}
\Edec>\Ktech\sqrt{\frac{\left(\log n\right)}{\log \left(\frac{1}{\epsilon}\right)}}\frac{Rn}{2}\label{eq:whatToDiviveBynR}
\end{equation}
which is the result we are seeking to prove.
\end{IEEEproof}
The following corollary is immediate.
\begin{cor}
If a sequence of decoding schemes in which in the limit of large $n$
$P_{e}^{\text{blk}, n}<\half$, the average decoding energy, per decoded bit
(which we denote $E_{\text{dec,avg}}$) is bounded as:
\begin{equation}
E_{\text{dec,avg}
}> \Ktech\sqrt{\frac{\left(\log_{2}n\right)}{\log_{2}\left(\frac{1}{\epsilon}\right)}}. \label{corollaryOne}
\end{equation}
\end{cor}
\begin{IEEEproof}
The proof follows simply by dividing (\ref{eq:whatToDiviveBynR})
by $nR$, the number of bits such a code is responsible for decoding.
\end{IEEEproof}

\section{Serial Computation}\label{sec:SerialComp}

Our result in Section~\ref{sec:Key-Result-of} applies to decoders implemented entirely in parallel;
however, this does not necessarily reflect the state of modern decoder implementations.
Below we provide a modified version of the Thompson VLSI model that
allows for the source of the computation to be input serially and
the outputs to be computed serially.

In this modified model, we assume that the circuit computes a function
of $n$ inputs and $k$ outputs. However, instead of having $n$ input
nodes and $k$ output nodes, the circuit has $p$ input nodes and
$j$ output nodes. The computation terminates after a set $\tau$
number of clock cycles, and during the $\tau$ clock cycles, the inputs
to the computation may be input into the $p$ input nodes (where $p$
bits at most can be input during a single clock cycle), and the outputs
of the computation must appear in the output nodes during specified
clock cycles of the computation.

\begin{rem}\label{simpleBoundForTau} The number of clock cycles $\tau$ must at least be enough to
output all the bits. If there are $j$ output nodes and $k$ outputs
to the function being computed, then there must be at least $\frac{k}{j}$
clock cycles. If all the inputs into the computation are being used,
then there must also be at least $\frac{n}{p}$ clock cycles, though
it is technically possible for some functions to have inputs that
``don't matter'' so this is not a strict bound for all functions.
\end{rem}

Hence, a lower bound on the energy complexity for this computation
is:
\[
\Eproc\ge\tech\Ac\frac{k}{j}
\]
where $\Ac$ is the area of the circuit.

\begin{thm}\label{serialTheorem}Suppose there is a sequence codes
together with decoding schemes with rate $R$ and block length $n$
approaching infinity. We label the block error probability of
the length $n$ decoder as $P_{e}^{blk,n}$. Also suppose that the
number of output pins remains a constant $j$. Then either (a) $\lim_{n\rightarrow\infty}P_{e}^{blk,n}\ge \half$
or (b) there exists some $n_{0}$ such that for all $n$ greater than
$n_{0}$ 
\[
\Edec\ge\frac{\tech\lw^{2}R^2n}{j\log\left(\frac{1}{\epsilon}\right)}
\left(\log n-j\right)=\Omega\left(n\log n\right).
\]
\end{thm}

To prove this theorem, instead of dividing the circuit into subcircuits, we will
divide the computation conceptually in time, by dividing the computation into epochs. 
More precisely, consider dividing the computation outputs into chunks of size $m$
(with the exception of possibly one chunk if $m$ does not evenly
divide $k$), meaning that there are $\left\lceil \frac{k}{m}\right\rceil $
such chunks. Hence, the outputs, which can be labeled $\left(k_{1},k_{2},\ldots,k_{k}\right)$
can be divided into groups, or a collection of subvectors $\left(K_{1},K_{2},\ldots,K_{\left\lceil \frac{k}{m}\right\rceil }\right)$
in which $K_{1}=\left(k_{1},k_{2},\ldots k_{m}\right)$, $K_{2}=\left(k_{m+1},k_{m+2},\ldots k_{2m}\right)$
and so on, until $K_{\left\lceil \frac{k}{m}\right\rceil }=\left(k_{m\left\lfloor \frac{k}{m}\right\rfloor },k_{m\left\lfloor \frac{k}{m}\right\rfloor +1},\ldots,k_{k}\right)$.
\begin{defn} The set of clock cycles in the computation
in which the bits in $K_i$ are output is considered to be the \textsl{$i$th epoch}.
\end{defn}
In our analysis, we are interested in analyzing the decoding problem
for chunks of the output
as defined above for an $m$ that we will choose later for the convenience
of our theorem. We are also interested in another set of quantities:
the input bits injected into the circuit between the time when the
last of the bits in $K_{i}$ are output and the first of the bits
in $K_{i+1}$ bits are output. Label the collection of these bits
as $\left(N_{1},N_{2},\ldots,N_{\left\lceil \frac{k}{m}\right\rceil }\right)$.
Label the size of each of these of these subvectors as $\left(n_{1},n_{2},\ldots,n_{\left\lceil \frac{k}{m}\right\rceil }\right)$,
so that the number of bits injected before all of the bits in $K_{1}$
are computed is $n_{1}$, and the number of those injected after the
first $n_{1}$ bits are injected and until the clock cycle when the
last of the bits in $K_{2}$ are output is $n_{2}$, and so on. Let $s_i $
be the number of erasures that are injected into the circuit during the
$i$th epoch. Note
that by Lemma~\ref{pidgeonholeLemma} an error occurs when 
\[
m\le n_{i}+\bar{A}-s_i
\]
where $\bar{A}=\frac{\Ac}{\lw^{2}}$ is the maximum number of
bits that can be stored in the circuit, remembering that according
to the computation model the maximum number of bits that can be stored
in a circuit must be proportional to the area of the circuit, as each
wire in the circuit at any given time in the computation can hold
only the value $1$ or $0$.

\begin{IEEEproof} (of Theorem \ref{serialTheorem})
Suppose we divide the circuit into chunks each of size $\bar{A}+j$,
$j$ more than the normalized circuit area. Then, if all the bits
$n_{i}$ are erased, the probability that at least one of the bits
of $K_{i}$ is not decoded must at least be $\half$, because
there are simply not enough non-erased inputs for the circuit to infer
the $m$ bits it is responsible for decoding in that window of time.
Note that we choose $m=\bar{A}+j$ so that an error event occurs with
probability at least $\half$ when all the $n_{i}$ bits are
erased, because it is technically possible that in a clock cycle that
outputs the last of the bits of $K_{i}$, $j-1$ bits of $K_{i+1}$
are output. Then, the number of bits required to be computed for the
next chunk of outputs is at least $k_{i+1}-j+1$. Let the size of
each $K_{i}$ (except possibly $K_{\left\lceil \frac{k}{m}\right\rceil }$)
be $\bar{A}+j$. Similar to what
we did for in Section~\ref{sub:Bound-on-Block}, denote the event that all input bits
in $N_{i}$ are erased as $W_{i}$.
Thus:

\begin{align*}
P\left(\text{correct}\right) & =  P\left(\cap_{i=1}^{\left\lceil \frac{k}{m}\right\rceil }\bar{W}_{i}\right)P\left(\text{correct}|\cap_{i=1}^{\left\lceil \frac{k}{m}\right\rceil }\bar{W}_{i}\right)+P\left(\cup_{i=1}^{\left\lceil \frac{k}{m}\right\rceil }W_{i}\right)P\left(\text{correct}|\cup_{i=1}^{\left\lceil \frac{k}{m}\right\rceil }W_{i}\right)\\
 & \le  \prod_{i=1}^{\left\lfloor \frac{k}{m}\right\rfloor }\left(1-\epsilon^{n_{i}}\right)+\frac{1}{2}.
\end{align*}
The first term is simplified by recognizing the independence of erasure events in the channel
and the second term is simplified by the fact that, conditioned on the event that at least one subcircuit has
input nodes being all erasure symbols, Lemma \ref{pidgeonholeLemma} applies and at least one subcircuit
must make an error with probability at least $\half$.
Thus:
\begin{align}
P_{e}^{blk,n} & =  1-P\left(\text{correct}\right)\nonumber \\
 & \ge  1-\prod_{i=1}^{\left\lfloor \frac{k}{m}\right\rfloor }\left(1-\epsilon^{n_{i}}\right).\label{eq:LowerBoundEquation}
\end{align}
It must be that $\sum_{i=1}^{\left\lceil \frac{k}{m}\right\rceil }n_{i}=n$,
and thus $\sum_{i=1}^{\left\lfloor \frac{k}{m}\right\rfloor }n_{i}\le n$,
where again $n$ is the total number of inputs.

We can apply Lemma~\ref{ConvexOptimizationLemma} to show that the product term in (\ref{eq:LowerBoundEquation}) is maximized when
each $n_{i}$ is equal to $n_{i}=\frac{n}{\left\lfloor \frac{k}{m}\right\rfloor }$. Thus, we show that:
\[
P_{e}^{blk}\ge1-\left(1-\epsilon^{\frac{n}{\left\lfloor \frac{k}{m}\right\rfloor }}\right)^{\left\lfloor \frac{k}{m}\right\rfloor }.
\]
For the sake of the convenience of calculation, we replace $\left\lfloor \frac{k}{m}\right\rfloor $
with $\frac{k}{m}$, which will not alter the evaluation of the limit by Lemma~\ref{roundingErrorLemma}, giving us:
\begin{equation}
P_{e}^{blk,n}\ge1-\left(1-\epsilon^{\frac{mn}{k}}\right)^{\frac{k}{m}} \label{lineToSub}
\end{equation}
Since we have assumed $m=\bar{A}+j$, suppose that $\bar{A}\le\frac{cR}{\log\frac{1}{\epsilon}}\log n-j$,
and recognizing that $k=Rn$, and that $m=\bar{A}+j$, substituting into (\ref{lineToSub}) and simplifying gives us:
\[
P_{e}^{blk}\ge\frac{1}{2}-\left(1-\exp\left(-c\log n\right)\right)^{\frac{\log\left(\frac{1}{\epsilon}\right)Rn}{\log n}}
\]
Thus, if $c<1$ and applying Lemma~\ref{limitLemma}:
\[
\lim_{n\rightarrow\infty}P_{e}^{blk}\ge\frac{1}{2}-\lim_{n\rightarrow\infty}\left(1-\exp\left(-c\log n\right)\right)^{\frac{\log\left(\frac{1}{\epsilon}\right)Rn}{\log n}}=\frac{1}{2}
\]
Hence, either in the limit block error probability is at least $\half$, or $\bar{A}>\frac{cR}{\log\frac{1}{\epsilon}}\log n-j$ and thus
\[
\Edec\ge\tech\lw^{2}\bar{A}\frac{k}{j}\ge\frac{\tech\lw^{2}R^2n}{j\log\left(\frac{1}{\epsilon}\right)}
\left(\log n-j\right)=\Omega\left(n\log n\right),
\]
where we have used the fact that the number of clock cycles is at
least $\frac{k}{j}$ as well as our bound on $\bar{A}$.
\end{IEEEproof}

\section{A General Case: Allowing the Number of Output Pins to Vary with Increasing Block Length} \label{generalCase}
The results in Sections \ref{sec:Key-Result-of} and \ref{sec:SerialComp} show that in the case of fully parallel implementations, the Area-Time complexity of decoders that asymptotically have a low block error probability must asymptotically have a super-linear lower bound. Technically, however, it may be possible to make a series of circuits with increasing block length, and have the number of output pins increase with increasing block length. We can show that, in this case, a super-linear lower bound exists as well where we require only weak assumptions on the circuit layout. This proof applies the main principle of this paper: namely that if a subcircuit has all its inputs erased, then that subcircuit must somehow have communicated to it other bits from outside this circuit, or it must, with high probability, make an error. In Theorem \ref{mainTheorem}, we recognize that in a fully parallel computation these bits must be injected to it from another part of the circuit, resulting in some energy cost. In Theorem \ref{serialTheorem}, we divide the circuit into epochs, and recognize that if all the input bits injected into the circuit during that epoch are erased then the circuit must have bits injected to it from before that epoch. But the number of bits that can be carried forward after each epoch is limited by the area of the circuit. In the general case in which the number of output pins can vary with block length, we divide the circuit into subcircuits and epochs (in essence, dividing the circuit in both time and space) and apply these two fundamental ideas.

To accomplish this lower bound, we will need this simplifying assumption: for any decoder with $j$ output pins, each output pin is responsible
for, before the end of the computation, outputting between $\left\lfloor \frac{k}{j}\right\rfloor $
and $\left\lceil \frac{k}{j}\right\rceil $ bits. Furthermore, we assume that each output bit produces an output at the same time. We call this assumption an \textsl{output regularity assumption}. This assumption allows us
to divide the circuit into subcircuits and then epochs, and thus with this assumption each subcircuit can be
divided into \textsl{subcircuit epochs}. The main structure of the proof will be this: if the energy of a computation is not high, then there will be many subcircuit epochs that do not have enough bits injected into
them to overcome the case of one of them having all of their input bits erased. The task is thus to choose a correct
number of divisions of the circuit into subcircuits and epochs, so that the probability of this event (that a subcircuit epoch makes an error) is high unless the area-time complexity of the computation is high.

\begin{thm} \label{generalTheorem} For a sequence of codes and circuit implementations
of decoding algorithms in which block length $n$ gets large, and
where the number of output pins $j$ can vary with block length $n$,
and the computation performed by the decoders is consistent with the output regularity assumption, then, in the limit as $n$ approaches
infinity, either (a) $\lim_{n\rightarrow\infty}P_{e}^{blk,n}\ge\frac{1}{2}$
or (b) for a sufficiently large $n$, $\Edec\ge\Omega\left(n\left(\log n\right)^{\fifth}\right)$.
\end{thm}
\begin{IEEEproof}
The proof is given in Appendix~\ref{AppendixGeneralProof}.
\end{IEEEproof}

\section{Consequences} \label{Sec:Consequences}

A direct consequence of our work is that as code rates approach capacity, the average
energy of decoding, per bit, must approach infinity. It is well known 
from \cite{Strassen} and \cite{GallagerBook} and further studied in \cite{Polyanskiy} that as a function of fraction of
 capacity $\eta=\frac{R}{C}$,
 the minimum block length scales approximately as
\[
n\approx\frac{c}{\left(1-\eta\right)^{2}}
\]
for a constant $c$ that depends on the channel and target probabilities of error. We are not concerned about
the value of this constant, but rather the dependence of this approximation on $\eta$. Plugging this result into (\ref{corollaryOne})
implies:
\[
E_{\text{dec,avg}}\gtrsim \Ktech\sqrt{\frac{\left(\log_{2}\frac{c}{\left(1-\eta\right)^{2}}\right)}{\log_{2}\left(\frac{1}{\epsilon}\right)}}=\Omega\left(\sqrt{\log\left(\frac{1}{1-\eta}\right)}\right).
\]
This result implies that not only must the total energy of a decoding algorithm
approach infinity as capacity is approached (this is a trivial consequence
of the fact that block length must approach infinity as capacity is approached), 
but also the energy \emph{per bit} must approach infinity as capacity is approached.
Thus, if the total energy per bit is to be optimized, a rate strictly less than capacity must be
used. We cannot get arbitrarily close to optimal energy per bit by getting arbitrarily close to capacity,
which would be the case if there were linear energy complexity algorithms with block error
probability that stay less than $\half$.

The result of Theorem~\ref{serialTheorem} can also be extended to find a fundamental lower bound
on the average energy per bit of serially decoded, capacity-approaching codes. For the same reason
as in the fully parallel case, we can see that as a function of gap to capacity, the average energy
per bit for a decoder must scale as
\[
E_{\text{dec,avg}}\ge\frac{2\lw^{2}R}{j\log\left(\frac{1}{\epsilon}\right)}\left(\log\left(\frac{1}{1-\eta}\right)-j\right)=\Omega\left(\log\left(\frac{1}{1-\eta}\right)\right).
\]
Finally, it can be shown from Theorem~\ref{generalTheorem} that in circuits in which the output pins can grown arbitrarily, and the regular output rate condition is satisfied, the average energy per bit as a function of gap to capacity must scale as
\[
E_{\text{dec,avg}}\ge\Omega\left(\log\left(\frac{1}{1-\eta}\right)^{\fifth}\right).
\]
\section{Upper Bound on Energy of Regular LDPC Code\label{sec:upperBound}}

We have shown that for any code and decoding circuit with block error
probability that is below $\half$, the Area-Time complexity must scale
at least as fast as $\Omega\left(n\sqrt{\log n}\right)$. We
provide here an example of a particular circuit layout that achieves
$O\left(n^{2}\right)$ complexity. Low density parity check (LDPC) codes are standard codes first
described by Gallager in \cite{GallagerLDPC}. There have been a number
of papers that have sought to find very energy-efficient implementations
of LDPC decoders; for example \cite{Darahiba}. The reference \cite{RothEtAl} gives an overview of various techniques used to create actual VLSI implementations of LDPC decoders. However,
these papers have not sought to view how the energy per bit of these
decoders scales with block length; they show a method to optimize
an LDPC decoder of a particular block length and show that their implementation
method improves over a previous implementation. Our goal is to provide
an understanding of how a particular implementation of LDPC codes
should scale with block length $n$.

We provide a simple circuit placement algorithm that results in a
circuit whose area scales no faster than $O\left(\left|E\right|^{2}\right)$
where $\left|E\right|$ is the number of edges in the circuit. For
a regular LDPC code with constant node degrees, this implies that
the area scales as $O\left(n^{2}\right)$.

\begin{figure} 
\centering
\includegraphics{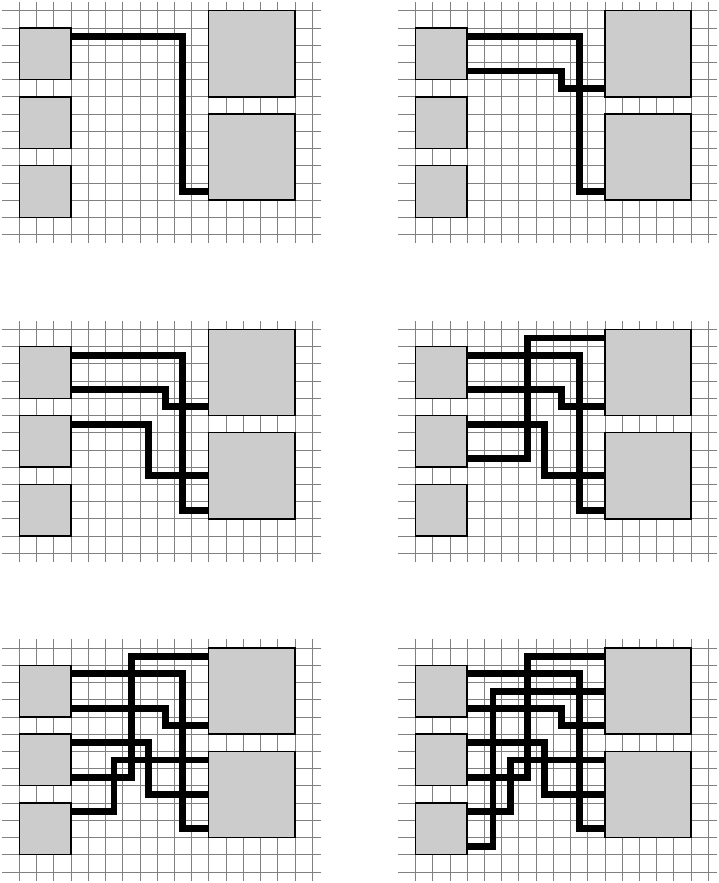}
\caption{An example of an implementation of an LDPC code Tanner graph. There are $\left|E\right|$ edges that correspond to interconnections that must be made. Going left to right, starting at the top left circuit, the six parts of this diagram show the progressive addition of each additional ``edge'' in the circuit implementation of the Tanner Graph. Each wire has a unique column that it is allowed to run along, and each output has a unique row, ensuring that no two wires ever need to run along the same section. The only time when the wires need to intersect is during a wire crossing, which is explicitly allowed by our circuit axioms. This method can be used to draw any arbitrary bipartite graph with $\left|E\right|$ edges.  } \label{LDPCNsquaresExample}
\end{figure}

The placement algorithm proposed involves actually instantiating the
Tanner Graph of the LDPC code with wires, where each edge of the Tanner
graph corresponds to a wire connected to $n$ subcircuits that perform variable node computations
and the $n-k$ subcircuits that perform check node computations. Our concern is
not about the implementation of the variable and check nodes in this
circuit. In the diagram, we treat these as merely a ``black box''
whose area is no greater than proportional to the square of the degree of the node. Of course, the
actual area of these nodes is implementation specific, but the important point is that the area of
each node should only depend on the particular node degree and not on the
block length of the entire code. Our concern is actually regarding how the area of the
interconnecting wires scales.
The wires leading out of each of these check and variable node subcircuits
correspond to edges that leave the corresponding check or variable
node of the Tanner graph. The challenge is then to connect the variable
nodes with the check nodes with wires as they are connected in the
Tanner graph in a way consistent with our circuit axioms. We lay out
all the variable nodes on the left side of the circuit, and all the
subcircuits corresponding to a check node on the right side of the
circuit, and place the outputs of each of these subcircuits in a unique
row of the circuit grid (see Fig.~\ref{LDPCNsquaresExample}). Note that the number
of outputs for each variable and check node subcircuit will be equal to
the degree of that corresponding node in the Tanner graph of the code.
 The height of this alignment of nodes will be $2\left|E\right|$,
twice the number of edges in the corresponding Tanner graph (as there must be a unique row
for each of the $\left|E\right|$ edges of the variable nodes and also for the $\left|E\right|$
edges leading from the check nodes. 

The distance between these columns of check and variable nodes is
$\left|E\right|$. Each output of the variable nodes is assigned a
unique grid column that will not be occupied by any other wire (except
in the case of a crossing, which according to our model is allowed).
A horizontal wire is drawn until this column is reached, and then
the wire is drawn up or down along this column until it reaches the
row corresponding to the variable node to which it is to be attached. A
diagram of the procedure to draw such a circuit for a case of $6$
edges is shown in Fig.~\ref{LDPCNsquaresExample}. Since each output
of the variable and check node ``black boxes'' takes up a unique
row, and each wire has a unique column, no two wires in drawing this
circuit can ever run along the same edge; they can only cross, which
is permitted in our model.

The total area of this circuit is thus bounded by: $\Ac \le \An+\Aw$,
where $\An$ is the area of the nodes and $\Aw$ is the area of the wires. Now it is sufficient that there is a grid row for each output of the
variable nodes and the check nodes, and that there is a column for
each edge. Hence 
\[
\Aw \le2\left|E\right|\cdot\left|E\right|=2\left|E\right|^{2}.
\]

We assume that the area of the subcircuits that perform the computational
node operations can complete their operation in one clock cycle and
take up area proportional to the square of their degree. Hence we suppose that $\An\le d_{v}^2n+d_{c}^2\left(n-k\right)$,
where $d_{v}$ is the degree of the variable nodes and $d_{c}$ the
degree of the check nodes. We then conclude that:
\[
\Ac \le2\left|E\right|^{2}+d_{v}^2n+d_{c}^2\left(n-k\right).
\]

The total energy for the computation will depend on the number of iterations
performed. Since each iteration requires sending information for the
variables nodes to the check nodes and back again, this can be performed
in $2$ clock cycles. Hence, $\tau=2N$, where $N$ is the number
of iterations performed, and of course $\tau$ is the number of clock
cycles in the computation.

Thus, the total energy of this implementation of an LDPC code is
upper bounded by
\[
\Edec\le2N\left(2\left|E\right|^{2}+d_{v}^2n+d_{c}^2\left(n-k\right)\right).
\]

The work of Lentmaier \emph{et al.} \cite{1522644} has shown that for an LDPC decoder, for asymptotically low block error
probability, $\tau=O\left(\log\log n\right)$ iterations are sufficient if the node degrees are high enough. This
then results in an upper bound on the energy of 
\[
\Edec\le2N\left(2\left(nd_{v}\right)^{2}+d_{v}^2n+d_{c}^2\left(n-k\right)\right)=O\left(n^{2}\log\log n\right).
\]

\section{Conclusion}

This work expands on previous work in \cite{groverFundamental}
by providing a standard to which decoding algorithms together with circuit
implementations can be compared. Earlier work on the energy used
in decoding (for example, \cite{Darahiba}) have involved
trying to optimize a circuit that implements a particular code; they
have not sought to understand how the energy scales with the length
of the code.

Some work has provided an
analysis of the energy requirements for specific types of codes. The work in \cite{korb} has provided
a way to analyze the energy requirements for an LDPC decoder.
The result of our paper
is more general: it applies to any decoding algorithm. Further investigations
should compare the results in this paper to existing results relating energy
per bit with parameters like block error probability.

Finally, this paper should also be used to guide the development of new codes that attempt to approach
this fundamental lower bound. There may be some modifications of some types of codes, for example
LDPC codes, whose Area-Time complexity
 is chosen to be $O\left(n\sqrt{\log n}\right)$ (for example, by choosing the neighbors
of the nodes in the Tanner graph representation of such a code to limit the area of the code, or by
limiting the number of iterations). Most analyses of LDPC codes assume a random Tanner graph. If
the interconnections of the LDPC code are restricted to limit the area of the implementation
(and thus violating the assumptions of most LDPC code
analyses) will the decoder still have a good performance? The work of \cite{1023755} suggests there is
some kind of trade-off.

\appendix

\renewcommand{\thesubsection}{\Alph{subsection}}

\subsection{Proof Of Theorem~\ref{generalTheorem}}\label{AppendixGeneralProof}

To accomplish this super-linear lower bound, we divide how the number
of output pins $j$ scales with $n$, the block length, into cases.
We suppose that $j\ge\sqrt{\log k}$. If not, using our result from
Theorem~\ref{serialTheorem}, for codes with asymptotic block error probability
less than $\half$, 
\[
\Edec\ge\frac{\lw^{2}R^2n}{\log\left(\frac{1}{\epsilon}\right)}\left(\frac{\log n}{j}-1\right),
\]
if $j<\sqrt{\log k}$ then we can show that 
\[
A\tau\ge\frac{\lw^{2}R^2n}{\log\left(\frac{1}{\epsilon}\right)}\left(\sqrt{\log k}-1\right)=\Omega\left(n\sqrt{\log n}\right)\ge\Omega\left(n\left(\log n\right)^{\fifth}\right)
\]
and we are done. 

\begin{rem}\label{divideIntoSeparateSubsequencesRemark} Technically, the statement that either $j\ge\sqrt{\log k}$
or $j<\sqrt{\log k}$ does not fully specify all possible sequences
of output pins. However, for any sequence, we can divide the sequences
into separate subsequences, specifically the sequences of codes in
which $j\ge\sqrt{\log k}$ and in which $j<\sqrt{\log k}$. For each
of those subsequences we can prove our lower bound. 
\end{rem}

Let $\bar{A}=\frac{A}{\lambda^{2}}$ be the normalized circuit area.
Suppose also that $\frac{\bar{A}}{j}\le\log^{0.9}n$. Otherwise, if
$\frac{\bar{A}}{j}>\log^{0.9}n$ and from our simple bound in Remark
\ref{simpleBoundForTau}, we can see that
\[
\bar{A}\tau\ge\bar{A}\frac{k}{j}>k\log^{0.9}n\ge\Omega\left(n\left(\log n\right)^{\fifth}\right)
\]
and we are done.

Note again that, just as in Remark \ref{divideIntoSeparateSubsequencesRemark}, if the area alternates
between $\log^{0.9}n$ with increasing block length, we can simply
divide the sequence of decoders into two subsequences and prove that
the necessary scaling law holds for each subsequence.

Hence, we consider the case that we have a sequence
of serial decoding algorithms in which the area of the circuit grows
with the block length $n$ and the number of output nodes on the circuit
grows with $n$. We consider the case in which 
\begin{equation}
\frac{\bar{A}}{j}\le\log^{0.9}n \label{areaScalingAssumption}
\end{equation}
 and 
\begin{equation}
j\ge\sqrt{\log k}
\end{equation}

We will now choose a way to divide the computation into $M$ epochs
and $N$ subcircuits. 

For each of the $M$ epochs we want the number of bits responsible
on average for each decoder to decode to be four times the area. This
will mean that, even if we optimistically assume that before the beginning
of each epoch a circuit had already computed the future outputs, a
typical subcircuit can only store a fraction of the bits it is responsible
for decoding in the next epoch. Note that the number of bits that
a subcircuit is responsible for in total over the entire computation
must be $\frac{k}{N}$ and hence, if the computation is to be divided
into $M$ epochs, during each epoch, an average subcircuit must be
responsible for decoding $\frac{k}{MN}$ bits. We seek to choose an
$M$ such that
\[
\frac{k}{MN}\ge4\bar{A}_{\text{subckt,avg}}
\]
where $\bar{A}_{\text{subckt,avg}}$ is the average normalized area of a subcircuit. This will be true if 
$
\frac{k}{MN}\ge4\frac{\bar{A}}{N}
$
or equivalently if
$
M\le\frac{k}{4\bar{A}}
$
so we choose 
\[
M=\frac{k}{4\bar{A}}
\]
We also want 
$
NM=\frac{cn}{\log n}
$
for a constant $c$ which we will choose later, so we choose
\[
N=\frac{cn4\bar{A}}{k\log n}=\frac{c4\bar{A}}{R\log n}.
\]

We need to show that this is a valid choice for $N$. The restriction
on the choice of $N$ is that $N\le j$ (we can't subdivide the circuit
into more subcircuits than there are output pins). By applying the assumption on the scaling of the area of the circuit in (\ref{areaScalingAssumption}) we can see that
\[
\frac{4\bar{A}}{R\log n}\le\frac{4cj\log^{0.9}n}{R\log n}=j\left(\frac{4c}{R}\right)\frac{\log^{0.9}n}{\log n}
\]
is asymptotically less than $j$, and hence this choice of $N$ is
valid. Our choice of $M$ is $\frac{k}{4A}$. The restriction on the
choice of $M$ is that $M\le\frac{k}{j}$ (there must be at least
one output per pin per epoch). Thus $\frac{k}{4\bar{A}}\le\frac{k}{j}$, which
will be true when $j\le4A$. But since the $j$ output pins form part
of the area of the circuit, this must always be satisfied.

On a minor technical note, we can only choose integer values of $M$. Hence, we can decide to choose the floor of $M$. But,
as argued in Lemma~\ref{roundingErrorLemma} if the function for choosing $M$ grows with $n$ then the evaluation of a limit where we neglect this floor function is the same. So, our other requirement,
that $\lim_{n\rightarrow\infty}\frac{\bar{A}}{n}=0$ means that our
choice for $M$ grows as $n$ increases. We consider the case when area of the computation
remains proportional to $n$ at the end of this section.

Let the number of bits injected into the $i$th subcircuit during
the $q$th epoch from other parts of the circuit be $b_{i,q}$. Now, the average subcircuit has area
$\frac{\bar{A}}{N}$, so we consider the set of subcircuits that have area
less than $2\frac{\bar{A}}{N}$. This number must be at least $\half$
the subcircuits, otherwise the total area of all these subcircuits
would exceed the total circuit area. Denote the set of indices of
subcircuits with area less than $2\frac{\bar{A}}{N}$ as $T$. Note that
$\left|T\right|\ge\frac{N}{2}$.

Consider a specific $q$th epoch. Suppose that the total number of
bits injected into the subcircuits indexed by $T$ after $r$-stages
of nested bisections during this epoch is less than $\frac{j}{2}$.
If this is true, then there must be at least half of the subcircuits
in $T$ that have fewer than $\frac{2j}{N}$ bits injected into them.
Otherwise, the total number of bits injected into these subcircuits
is at least $\frac{\left|T\right|}{2}\frac{2j}{N}\ge\frac{N}{4}\frac{2j}{N}=\frac{j}{2}$,
which we assumed is not the case. 

Thus, with our assumptions, at the $q$th epoch, either the total
number of bits injected across $r$-stages of nested minimum bisections is
at least $\frac{j}{2}$, or there are at least $\frac{N}{4}$ subcircuits
with area less than $2\bar{A}$ that have less than $\frac{j}{2N}$ bits
injected into them. Denote the set of indices of these low area subcircuits
with a low number of bits injected into them during the $q$th epoch
as $S_{q}$. The size of $S_{q}$ we have assumed to be at least $\frac{N}{4}$,
so for the sake of simplicity define $S_{q}^{*}$ as a subset of $S_{q}$
with size exactly $\frac{N}{4}$. Now, consider the number of epochs
during which there are less than $\frac{j}{2}$ bits injected across
all the bisections. Either this number is less than $\frac{M}{2}$
or greater than or equal to $\frac{M}{2}$. Suppose that it is greater
than or equal to $\frac{M}{2}$. Denote the set of indices denoting
the epochs in which the number of bits injected across all the bisections
during that epoch is less than $\frac{j}{2}$ as $Q$, and a particular
set of size exactly $\frac{M}{2}$ as $Q^{*}$ (chosen for simplicity
of computation). 

We now apply the key principle used for all the theorems in this paper.
Consider a particular $q$th epoch where $q\in Q^{*}$, an epoch with
less than $\frac{j}{2}$ bits communicated across all the bisections.
During this epoch the subcircuits in $S_{q}$ are those with less
than $\frac{2j}{N}$ bits injected into them, and they have area at
most $\frac{2\bar{A}}{N}$, and are responsible for decoding $\frac{4\bar{A}}{N}$
bits. Let the number of input bits injected into the circuit for such
a particular subcircuit be $n_{i,q}$. If all of these inputs are
erased, then, by applying Lemma~\ref{pidgeonholeLemma} the circuit must guess at least $1$ output, and the
probability of error is at least $\half$, because in this case
they only have $B_{i,q}+\frac{2\bar{A}}{N}\le\frac{2j}{N}+\frac{2\bar{A}}{n}\le\frac{4\bar{A}}{N}$
bits to use. Using the same argument as in Theorems~\ref{serialTheorem} and \ref{mainTheorem},
we can show that if, for all the subcircuits in $S_{q}^{*}$, where
$q\in Q^{*}$, in the event that for any of these subcircuits all
of their $n_{i,q}$ input bits are erased, then an error occurs with
probability $\half$. Applying this principle gives us:

\begin{equation}
P_{e}^{blk}\ge\frac{1}{2}-\prod_{q=Q^{*}}\prod_{i\in S_{q}^{*}}\left(1-\epsilon^{n_{i,q}}\right)\label{eq:equationToConvexOptimize}
\end{equation}
where we note as well that 
\[
\sum_{q=1}^{\frac{M}{2}}\sum_{i=1}^{\frac{N}{4}}n_{i,q}\le n.
\]

Subject to
those restrictions, Lemma~\ref{ConvexOptimizationLemma} implies that the expression in (\ref{eq:equationToConvexOptimize})
is minimized when each of the $n_{i,q}$ are equal to $\frac{8n}{NM}$.
Hence
\begin{align*}
P_{e,blk} & \ge  \frac{1}{2}-\left(1-\epsilon^{\frac{8n}{NM}}\right)^{\frac{NM}{8}}\\
 & \ge  \frac{1}{2}-\left(1-\epsilon^{8c\log n}\right)^{\frac{\log n}{8n}},
\end{align*}
which, by applying Lemma~\ref{limitLemma}, can easily be shown to approach $\half$
when $n$ gets larger, if $c$ is chosen to be $\frac{\log\frac{1}{\epsilon}}{8}$.
Hence, either in the limit block error probability approaches $\half$
or the size of $Q$ is greater than $\frac{M}{2}$, and there are
many bits communicated in the circuit for many epochs.

From
Lemma~\ref{lem:EdecLemma}, by recognizing that for the circuit under consideration there are at least $j$ nodes, if there are
$B_r$ bits injected across all the $r$-stages of nested minimum bisections, then
\[
\bar{A}\tau\ge K' B_{r}\sqrt{\frac{j}{2^{r}}}
\]
where $K'=\frac{\sqrt{2}-1}{4\sqrt{2}}$ and $2^{r}=N$, the number of subcircuits into which the circuit was divided. Thus, combining this bound with our choice for $N$ and the
assumption that there are at least $\frac{j}{2}$ bits injected across
all the bisections for epochs in $Q^{*}$, we get that either
\[
\bar{A}\tau_{i}\ge K'\frac{j}{2}\sqrt{\frac{j}{N}}
\]
for at least $\frac{M}{2}$ epochs, or $\lim_{n\rightarrow\infty P_{e}^{blk,n}}\ge\frac{1}{2}$.
Hence, in total,

\begin{align}
\bar{A}\tau & \ge  K'\frac{j}{2}\sqrt{\frac{j}{N}}\frac{M}{2}\nonumber \\
 & =  K'\frac{j}{2}\sqrt{\frac{j}{N}}\frac{k}{8\bar{A}}\nonumber \\
 & =  K'\frac{j^{1.5}}{2}\sqrt{\frac{R\log n}{4c\bar{A}}}\frac{k}{8\bar{A}} \nonumber\\
\bar{A}^{2.5}\tau & \ge  \frac{K'}{32}kj^{1.5}\sqrt{\frac{R\log n}{c}}\label{eq:ApplyingAssumptionOfHowJGrows-1}
\end{align}

We also have the bound from Remark~\ref{simpleBoundForTau}:
\begin{align*}
\tau & \ge  \frac{k}{j}\text{, which implies}\\
\tau^{1.5} & \ge  \frac{k^{1.5}}{j^{1.5}}
\end{align*}
and hence, combining this with (\ref{eq:ApplyingAssumptionOfHowJGrows-1}), we get
\begin{align*}
\bar{A}^{2.5}\tau^{2.5} & \ge \frac{K'}{32}k^{2.5}\sqrt{\frac{R\log n}{c}}\\
\bar{A}\tau & \ge \frac{\left(K^{'}\right)^{\frac{2}{5}}}{4}k\left(\frac{8R\log n}{\log\left(\frac{1}{\epsilon}\right)}\right)^{\frac{1}{5}}=\Omega\left(k\left(\log n\right)^{\frac{1}{5}}\right).
\end{align*}

Finally, we must consider a case when the area of the circuit scales with $n$. This must
be treated separately because in this case our choice for $M$ in the above argument does not necessarily
grow with $n$ and so we can't assume that our rounding approximation is valid.
Thus, suppose that
$A=cn$. Suppose also that $j\le\frac{k}{\left(\log n\right)^{0.9}}$.
Then
\[
\tau\ge\frac{k}{j}\ge\left(\log n\right)^{0.9}
\]
from Remark~\ref{simpleBoundForTau}, and therefore the total Area-Time complexity of such
a sequence of decoders scales as 
\[
A\tau\ge cn\left(\log n\right)^{0.9}=\Omega\left(n\left(\log n\right)^{0.9}\right).
\]

In the other case, when $j\ge\frac{k}{\left(\log n\right)^{0.9}}$
then we can subdivide the circuit into $\frac{n}{\log n}$ pieces,
and make the same argument that has been made in Theorem~\ref{mainTheorem} that
the number of bits communicated across all cuts during the course
of the computation must be proportional to $k/2$. Recognizing that we have assumed there are at least
$cn$ nodes in the circuit and applying Lemma~\ref{lem:EdecLemma}, and also substituting $2^{r}=\frac{\log\frac{1}{\epsilon}n}{\log n}$
we get:
\[
\bar{A}\tau\ge\frac{\sqrt{2}-1}{8\sqrt{2}}\sqrt{\frac{cR\log n}{\log\left(\frac{1}{\epsilon}\right)}}n=\Omega\left(n\sqrt{\log n}\right)
\]
which of course is asymptotically faster than $\Omega\left(n\left(\log n\right)^{\fifth}\right)$.


\begin{thebibliography}{10}
\providecommand{\url}[1]{#1}
\csname url@samestyle\endcsname
\providecommand{\newblock}{\relax}
\providecommand{\bibinfo}[2]{#2}
\providecommand{\BIBentrySTDinterwordspacing}{\spaceskip=0pt\relax}
\providecommand{\BIBentryALTinterwordstretchfactor}{4}
\providecommand{\BIBentryALTinterwordspacing}{\spaceskip=\fontdimen2\font plus
\BIBentryALTinterwordstretchfactor\fontdimen3\font minus
  \fontdimen4\font\relax}
\providecommand{\BIBforeignlanguage}[2]{{%
\expandafter\ifx\csname l@#1\endcsname\relax
\typeout{** WARNING: IEEEtran.bst: No hyphenation pattern has been}%
\typeout{** loaded for the language `#1'. Using the pattern for}%
\typeout{** the default language instead.}%
\else
\language=\csname l@#1\endcsname
\fi
#2}}
\providecommand{\BIBdecl}{\relax}
\BIBdecl

\bibitem{Shannon}
C.~E. Shannon, ``A mathematical theory of communication,'' \emph{Bell Sys.
  Techn. J.}, vol.~27, no.~3, pp. 379--423 \& 623--656, 1948.

\bibitem{ElGamal}
A.~El~Gamal, J.~Greene, and K.~Pang, ``{VLSI} complexity of coding,'' \emph{The
  MIT Conf. on Adv. Research in {VLSI}}, 1984.

\bibitem{groverFundamental}
P.~Grover, A.~Goldsmith, and A.~Sahai, ``Fundamental limits on the power
  consumption of encoding and decoding,'' in \emph{Proc. 2012 IEEE Int. Symp.
  Info. Theory}, 2012, pp. 2716--2720.

\bibitem{ThompsonThesis}
C.~D. Thompson, ``A complexity theory for {VLSI},'' Ph.D. Thesis,
  Carnegie-Mellon, 1980.

\bibitem{knuth1976big}
D.~E. Knuth, ``Big omicron and big omega and big theta,'' \emph{ACM SIGACT
  News}, vol.~8, no.~2, 1976.

\bibitem{Thompson}
C.~D. Thompson, ``Area-time complexity for {VLSI},'' \emph{Proc. 11th Ann. ACM
  Symp. Theory of Comput.}, pp. 81--–88, 1979.

\bibitem{Howard}
S.~L. Howard, C.~Schlegel, and K.~Iniewski, ``Error control coding in low-power
  wireless sensor networks: When is {ECC} energy-efficient?'' \emph{EURASIP J.
  on Wireless Commun. and Netw.}, pp. 1--14, 2006.

\bibitem{Rabaey}
J.~Rabaey, A.~Chandrakasan, and B.~Nikolic, \emph{Digital Integrated
  Circuits}.\hskip 1em plus 0.5em minus 0.4em\relax Englewood Cliffs, NJ, USA:
  Prentice Hall, 2003.

\bibitem{WeiYuBenSmith}
W.~Yu, M.~Ardakani, B.~Smith, and F.~R. Kschischang, ``Complexity-optimized
  low-density parity-check codes for {G}allager decoding algorithm {B},'' in
  \emph{Proc. 2005 IEEE Int. Symp. on Info. Theory}, 2005, pp. 1488--1492.

\bibitem{CooleyTukey}
J.~Cooley and J.~W. Tukey, ``An algorithm for the machine calculation of
  complex {F}ourier series,'' \emph{Math. Comput.}, 1965.

\bibitem{1023755}
J.~Thorpe, ``Design of {LDPC} graphs for hardware implementation,'' in
  \emph{Proceedings of 2002 IEEE International Symposium on Information
  Theory}, 2002, p. 483.

\bibitem{876069}
C.-H. Yeh, E.~Varvarigos, and B.~Parhami, ``Multilayer {VLSI} layout for
  interconnection networks,'' in \emph{Proc. Int’l Conf. Parallel
  Processing}, 2000, pp. 33--40.

\bibitem{Denker}
J.~S. Denker, ``A review of adiabatic computing,'' in \emph{1994 {IEEE}
  Symposium on Low Power Electronics}, 1994, pp. 94--97.

\bibitem{GroverInfoFriction}
\BIBentryALTinterwordspacing
P.~Grover, ``'{I}nformation-friction' and its implications on minimum energy
  required for communication,'' \emph{CoRR}, vol. abs/1401.1059, 2014.
  [Online]. Available: \url{http://arxiv.org/abs/1401.1059}
\BIBentrySTDinterwordspacing

\bibitem{Garey1976237}
\BIBentryALTinterwordspacing
M.~Garey, D.~Johnson, and L.~Stockmeyer, ``Some simplified {NP}-complete graph
  problems,'' \emph{Theoretical Computer Science}, vol.~1, no.~3, pp. 237 --
  267, 1976. [Online]. Available:
  \url{http://www.sciencedirect.com/science/article/pii/0304397576900591}
\BIBentrySTDinterwordspacing

\bibitem{Strassen}
V.~Strassen, ``Asymptotische absch\"atzungen in {S}hannon’s
  {I}nformationstheorie,'' in \emph{in Transactions of the 3rd Prague
  Conference on Information Theory, Statistical Decision Functions, Random
  Processes}.\hskip 1em plus 0.5em minus 0.4em\relax Prague: Pub. House of the
  Czechoslovak Academy of Sciences, 1962, pp. 689--723.

\bibitem{GallagerBook}
R.~G. Gallager, \emph{Information Theory and Reliable Communication}.\hskip 1em
  plus 0.5em minus 0.4em\relax New York, NY, USA: John Wiley \& Sons, Inc.,
  1968.

\bibitem{Polyanskiy}
Y.~Polyanskiy, ``Channel coding: non-asymptotic fundamental limits,'' Ph.D.
  dissertation, Princeton University, Nov. 2010.

\bibitem{GallagerLDPC}
R.~Gallager, ``Low-density parity-check codes,'' \emph{Information Theory, IRE
  Transactions on}, vol.~8, no.~1, pp. 21--28, 1962.

\bibitem{Darahiba}
A.~Darabiha, A.~Chan~Carusone, and F.~R. Kschischang, ``Power reduction
  techniques for {LDPC} decoders,'' \emph{IEEE J. of Solid-State Circuits},
  vol.~43, no.~8, pp. 1835--1845, Aug. 2008.

\bibitem{RothEtAl}
C.~Roth, A.~Cevrero, C.~Studer, Y.~Leblebici, and A.~Burg, ``Area, throughput,
  and energy-efficiency trade-offs in the {VLSI} implementation of {LDPC}
  decoders,'' in \emph{2011 IEEE International Symposium on Circuits and
  Systems (ISCAS)}, May 2011, pp. 1772--1775.

\bibitem{1522644}
M.~Lentmaier, D.~Truhachev, K.~Zigangirov, and D.~Costello, ``An analysis of
  the block error probability performance of iterative decoding,'' \emph{IEEE
  Transactions on Information Theory}, vol.~51, no.~11, pp. 3834--3855, Nov
  2005.

\bibitem{korb}
M.~Korb and T.~G. Noll, ``{LDPC} decoder area, timing, and energy models for
  early quantitative hardware cost estimates,'' in \emph{2010 Int.\ Symp.\
  System on Chip}, Sep. 2010, pp. 169--172.

\end{thebibliography}


\end{document}